\documentclass[prd,showpacs,superscriptaddress,twocolumn,floatfix,preprintnumbers,altaffilletter, letterpaper]{revtex4-1}

\usepackage{amsmath,amssymb}   

\usepackage{xcolor}
\usepackage{graphicx}  
\graphicspath{{.} {fig/}}

\usepackage[colorlinks=true]{hyperref}
\usepackage[T1]{fontenc}

\begin{document}

 \preprint{IUCAA-14/2016}
 \preprint{LIGO-P1600094}

\title{Transient Classification in LIGO data using Difference Boosting Neural Network}

\author{N Mukund} 
\email{nikhil@iucaa.in}
\affiliation{Inter-University Centre for Astronomy and Astrophysics (IUCAA), Post Bag 4, Ganeshkhind, Pune 411 007, India}
\author{S Abraham}\email{sheelu@iucaa.in}
\affiliation{Inter-University Centre for Astronomy and Astrophysics (IUCAA), Post Bag 4, Ganeshkhind, Pune 411 007, India}
\author{S Kandhasamy}\email{shivaraj@phy.olemiss.edu}
\affiliation{The University of Mississippi, University, Mississippi 38677, USA}
\author{S Mitra}\email{sanjit@iucaa.in}
\affiliation{Inter-University Centre for Astronomy and Astrophysics (IUCAA), Post Bag 4, Ganeshkhind, Pune 411 007, India}
\author{N S Philip}\email{nspp@associates.iucaa.in}
\affiliation{Department of Physics, St. Thomas College, Kozhencherry, Kerala  689641, India}
\date{\today}
\date{\today}

\begin{abstract}
Detection and classification of transients in data from gravitational wave detectors are crucial for efficient searches for true astrophysical events and identification of noise sources. We present a hybrid method for classification of short duration transients seen in gravitational wave data using both supervised and unsupervised machine learning techniques. To train the classifiers we use the relative wavelet energy and the corresponding entropy obtained by applying one-dimensional wavelet decomposition on the data. The prediction accuracy of the trained classifier on nine simulated classes of gravitational wave transients and also LIGO's sixth science run hardware injections are reported. Targeted searches for a couple of known classes of non-astrophysical signals in the first observational run of Advanced LIGO data are also presented. The ability to accurately identify transient classes using minimal training samples makes the proposed method a useful tool for LIGO detector characterization as well as searches for short duration gravitational wave signals.
\end{abstract}

\maketitle


Detection of short duration gravitational waves (GW) in LIGO data requires reliable identification and removal of noise transients produced by variety of non-astrophysical sources~\cite{GW150914,GW151226}. Noise transients present in the data reduces the reliability of a GW detection by increasing its false alarm probability. Mitigation of noise transients is a major challenge in searches for GW, specially for short duration events where the signal can be easily mimicked by non-astrophysical transients of varied origin.  These often have waveform morphology close to that of the targeted signal, thus making the differentiation even more difficult~\cite{abbott2016characterization}.  

With the advent of big data analysis, machine learning has emerged as a useful tool to handle huge volumes of data and to interpret meaningful results from them. In the past few decades, machine learning algorithms such as Artificial Neural Network (ANN)  \citep{1994VA.....38..273A,2013PhRvD..88f2003B}, Support Vector Machines \citep{DBLP:journals/ml/CortesV95,cc01a}, Random Forest \citep{ref1}, Gaussian Mixture Model  \cite{2015CQGra..32u5012P} etc. found many applications in astronomy and occasionally have been used for the study of noise artefacts in GW analysis. Since the visual inspection of individual events and their classification is time consuming and prone to errors, machine learning methods are more effective and reliable for the detection of hidden signatures of astrophysical GW in the data. 

We present a hybrid classifier that combines features from supervised and unsupervised  machine learning algorithms to do the transient classification.  Our classifier performs an unsupervised hierarchical clustering on the incoming data to identify possible groups and a supervised Bayesian \citep{Bayes01011763} classifier to do the final classification.  The classifier code uses features extracted from wavelet analysis of the data in a fast and efficient manner using GPU and MPI parallelization techniques, whereby, making it a good candidate for real-time burst trigger classification and detector characterization.  When used to predict the class labels for an input data, the classifier ranks the most likely classes each with an associated probability (confidence level) that may be used to set a threshold to discard unreliable predictions. This multiple class prediction is useful to identify borderline examples in the feature space. In our study, the classifier was first tested on simulated data consisting of astrophysical bursts along with commonly observed instrumental glitches and then on the LIGO sixth science run burst hardware injections. Targeted searches for specific glitch types seen in Advanced LIGO first observation data were also carried out and the results are reported. Recent methods like deep learning \cite{krizhevsky2012imagenet, zevin2016gravity} using convolutional neural networks require large number of training data and are computationally expensive. 
The fact that we are able to represent the transient classes with minimal features and fewer training data samples makes our method less susceptible to such issues and speeds up the training process, making it suitable for realtime applications.

\section{Transient Events in GW  Data \label{sec:Burst}}
Table~\ref{tab:glitches} lists the transients used in our analysis.
 Standard searches for compact binary coalescences use matched filtering as the base algorithm~\cite{SathyaSanjeev91}, while the burst searches primarily look for excess power in the data along with time coincidence to trigger a detection \cite{Klimenko:2004qh,Klimenko_CWB_2008}. Both these searches are followed by extensive sanity checks, where the auxiliary channels insensitive to astrophysical signals are inspected to rule out possible terrestrial coupling~\cite{abbott2016characterization}. Auxiliary channels are often in thousands and their coupling with the GW strain sensitive channel is seen to fluctuate in time due to the dynamic nature of the instrument. This often makes the auxiliary channel veto procedure a daunting task. Incorporating a machine learning based veto procedure to identify well known classes of non-astrophysical transients can help discern the trigger right at the strain channel and thus reduce false alarms.

\begin{figure*}[t]
\includegraphics[width=0.49\textwidth]{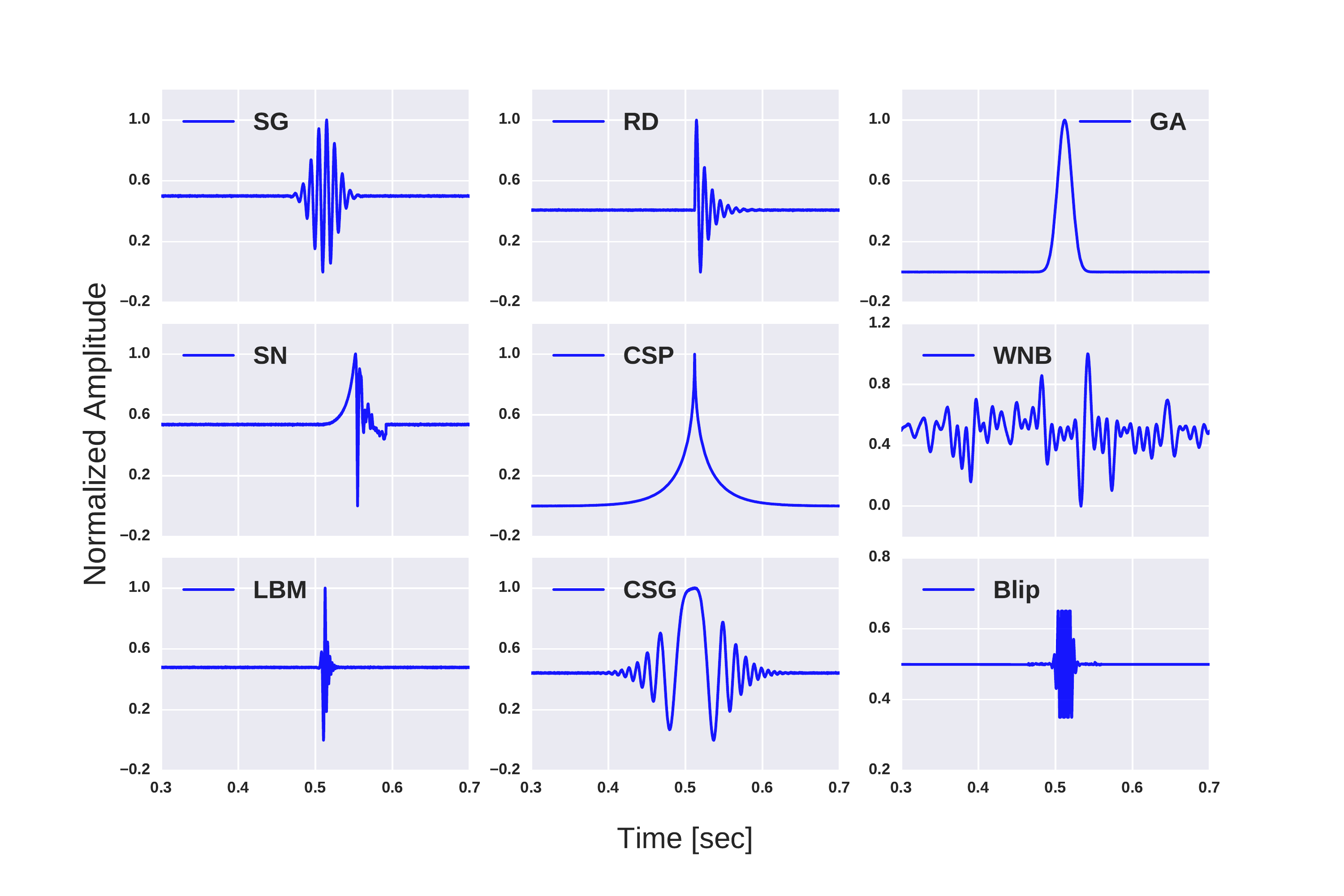}
\includegraphics[width=0.49\textwidth]{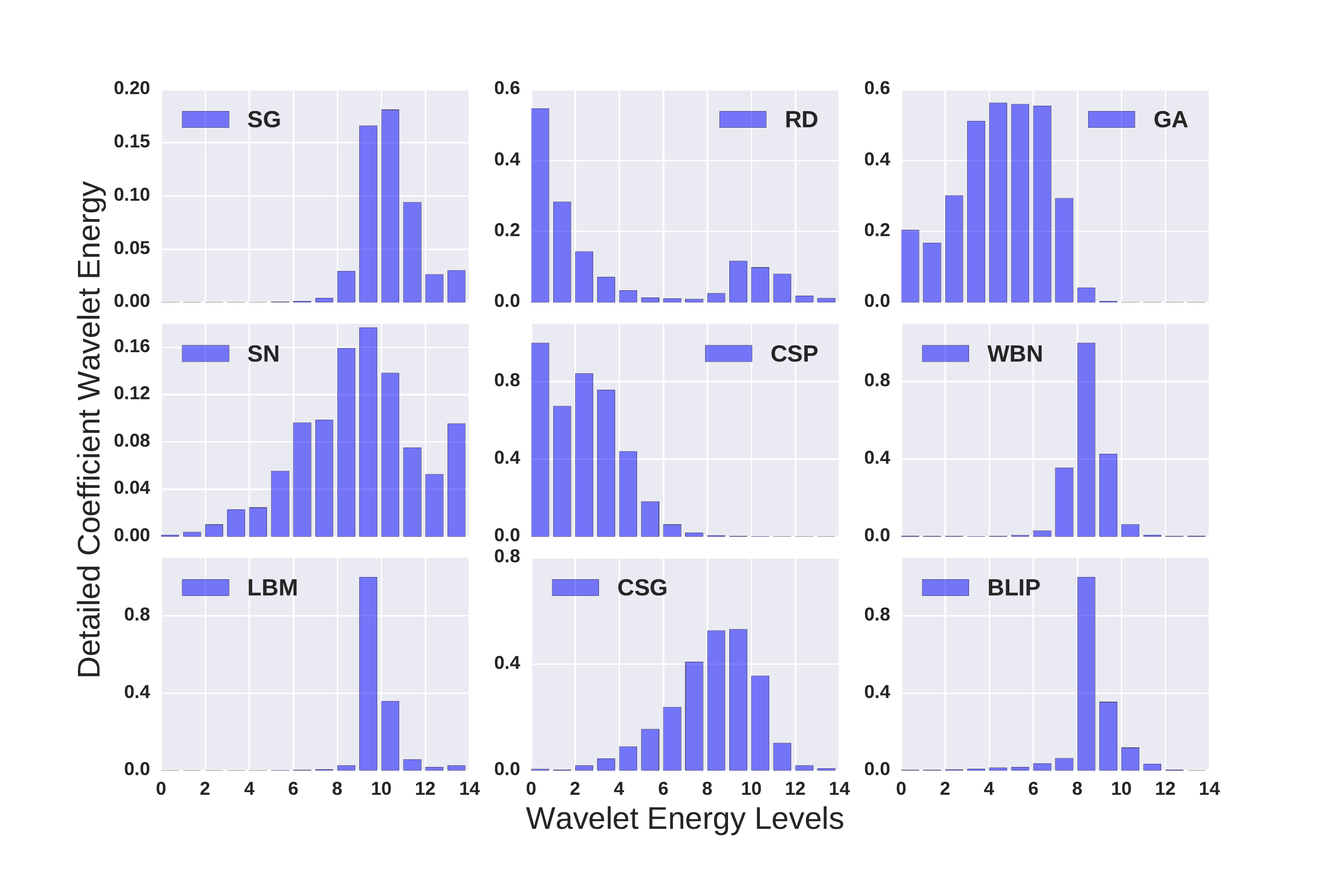}
\caption{Left panel depicts typical transient events (SNR set to 50 for better visualisation).  Wavelet energy median distribution for simulated data (SNR varied from 8 to 100)  shown in the right panel}\label{fig:ip}.
\end{figure*}

\section{\label{sec:DBNN} Classifier }

Machine learning involves techniques which allow systems to automatically learn and improve prediction accuracies by exploring their past experiences on data. It mimics human decision-making ability by discovering the relationships between the variables of a system from a given set of examples that have both the variables and the observed outcomes. Here we use a hybrid classifier, a supervised Bayesian~\cite{Bayes01011763} one called Difference Boosting Neural Network (DBNN)~\cite{Philip:2000:BDF:1294154.1294155,DBNN_Link}, to classify the burst signals.   

The DBNN can impose conditional independence on data without loss of accuracy in the posterior computations. It does this by associating a threshold window with each of the attribute values of the examples \citep{2002A&A...385.1119P}.  The network is designed to work with discrete value input features while GW data features are continuous. A simple method to deal with continuous feature value is to recast it into a suitable number of bins. There is no fixed criteria for the number of bins each feature may take. It might be argued that smaller the bin size, conditions can be imposed with better accuracy. However, in most practical situations, the optimal bin size is close to the square root of N, where N represents the number of discrete values present in the data for that variable. Once the bins are defined, for each feature bin and the given classes, the allowed ranges for all the remaining features are registered.
 
The DBNN, being a supervised neural network, requires a training data to configure the network before it can be used for classification of unseen data. The learning takes place by highlighting the difference between the features in two or more classes \cite{2002A&A...385.1119P} by using Bayesian probability as its central rule for decision making. The confidence in a prediction \cite{philip2010learning} is the value of the posterior Bayesian probability for a given set of input features. 

The working of DBNN can be divided into three units: Bayesian Probability Estimator, Gradient Descent Boosting Algorithm, and a Discriminant Function Estimator \cite{2002A&A...385.1119P}. The network starts with a flat prior for all the classes $ P(C_{k}) = 1/N $,  preventing the training from being biased to any specific prior distribution. The first unit in DBNN (executed by option 0 in the implementation) computes Bayesian probability and the threshold function for each of the training sample by constructing a grid for each class with columns representing the attributes and rows their respective values. The bin location for each attribute value is decided such that the full range of values can be uniformly covered by the set number of bins for that attribute across the classes. Initially the content in attribute bins are all set to one. The training examples are taken one by one and the bin corresponding to each attribute value for it's class is incremented by one. This sampled data is used to compute the likelihood for an attribute value to favour a class, $P(U_{m}|C_{k})$, as the ratio of occurrences (counts) in it's bin for the class $ C_{K} $ to the total counts in all $ k $ classes for the same bin number that $ U_{m} $ holds for that attribute. The classifier also makes notes for each attribute value and it's class, the allowed maximum and minimum values taken by the remaining attributes in the entire training sample. This information is used to negate the possibility that the value of one feature may favour multiple classes, unless all other features also have values in the same range across the classes. 

Though we started with a flat prior, to compute the Bayesian probability, we need to estimate the actual prior. In the Bayesian framework, prior has no special meaning. It is a weighted bias (belief) about the probable outcome of an experiment based on experiences in the past. In the second unit (executed by option 1 in the implementation), the DBNN estimates prior based on it's experience with the given training data. The DBNN does not make any change in the prior for correctly classified examples. In the case of failed examples, it attaches an additional weight to the attributes so that, it may also get correctly classified. To avoid random fluctuations due to the introduction of arbitrary priers, this is done by modifying the flat prior incrementally by  $\Delta W_{m} = \alpha(1 - \frac{P_{k}}{P_{k^{*}}})$ through a set of repeated rounds on the training data until the example gets correctly classified. That is, until $P(U|C_{k}) = \Pi_{m}P(U_{m}|C)$  goes to a maximum for the true class represented by the data. Here $P_{k}$ and $P_{k^{*}}$ respectively represent the calculated Bayesian probability for the true class and the wrongly estimated class and $\alpha$ is a fraction called the learning rate \cite{philip2010learning}. Since ratio of the probabilities are taken, this is much like the way humans arrive at their priers based on their cumulative experiences in the past. This process is called training, and after training, the estimated likelihoods and prior are saved for future use. The assumption during the training process is that a representative training data is available that has suitable examples to represent all the variants in the target space.

The third unit (executed by option 2 and 3 in the implementation) computes the discriminant function. According to Bayesian theorem, the updated belief or the posterior is the product of the prior and the evidence normalised over all possibilities. This can be written as 
\begin{equation}\label{eq:bay}
P(C_{k} \vert U) = \frac{\prod_{m}(P(U_m|C_k)W_m}{\sum_k \prod_m(P(U_m|C_k)W_m} 
\end{equation}
where W$_{m}$ represent the prior weight vector. 

DBNN has been successfully applied to many astronomical problems such as star-galaxy classification \cite{2002A&A...385.1119P}, classification of point sources such as quasars, stars and unresolved galaxies \cite{2012MNRAS.419...80A}, transient classification \cite{2012ASInC...6..151P} to indicate a few.

\begin{figure}[!htb]
\includegraphics[width=0.425\textwidth, height=0.415\textwidth]{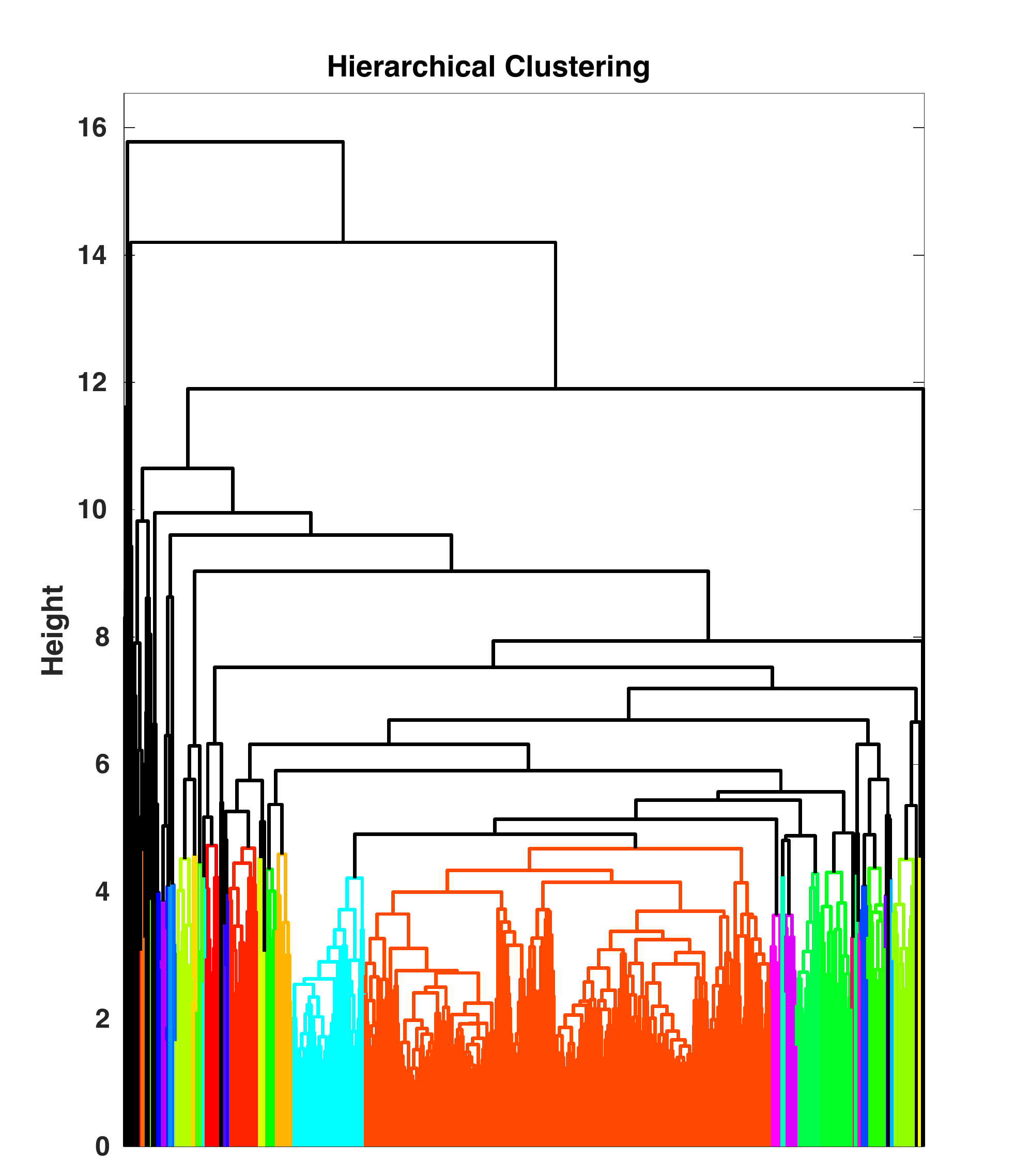}
\caption{Dendrogram showing hierarchical clustering of 1000 transient triggers identified in O1 Data from Hanford observatory by the {\tt Omicron} algorithm~\cite{OMICRON}.  The transient morphology changes progressively  from left to right}
\label{fig:Dendrogram}
\end{figure}

As for the case of all supervised networks, the accuracy of the predictions depend on the initial class selection and quality of the training data sets. When encountering real instrument data where it is difficult to know beforehand the actual groups present, running an unsupervised classifier prior to Wavelet-DBNN classifier was seen to vastly improve the results. This step becomes more relevant for targeted searches looking for a particular transient class where unsupervised learning can yield insights into contamination from other glitch classes.  Prior information about other glitches with very similar morphology can be made use of by the network to learn to differentiate between them whereby improving the accuracy. 

\begin{figure}[!htb]
\includegraphics[width=0.45\textwidth,scale=1.3]{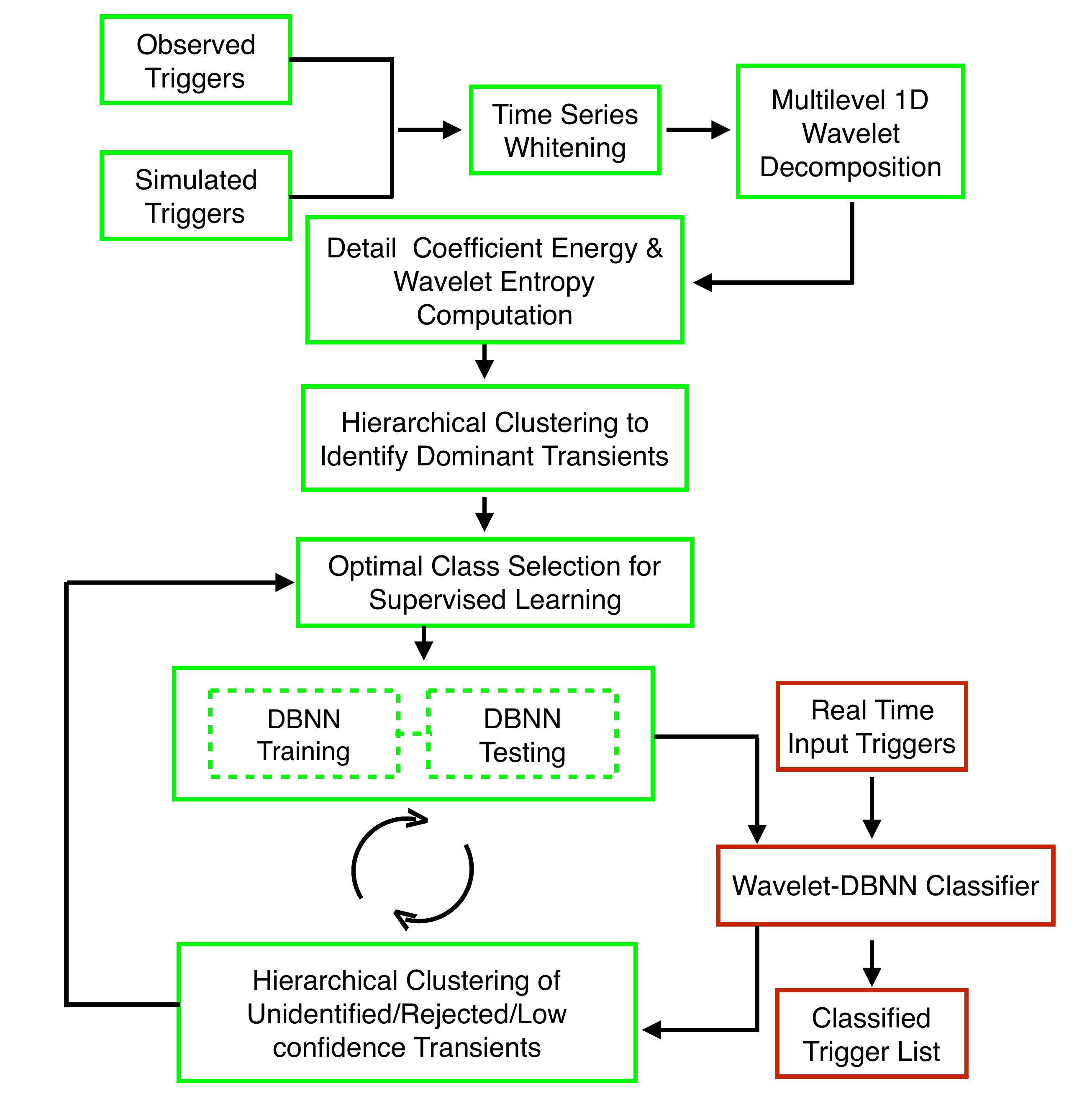}
\caption{Block diagram of the proposed hybrid classifier.} 
\label{fig:Schematic}
\end{figure}

We run an unsupervised classifier using Hierarchical clustering on the data to get an idea about the possible transient groups currently  present in the data and their respective distribution (see Figure \ref{fig:Dendrogram}). Classifier trained this way is observed to outperform the other scenarios where  class selection is done either by visual inspection  or by using predefined classes. 

We employ a bottom up agglomerate clustering where the pairwise distance is calculated using Mahalanobis distance measure \cite{MAHALANOBIS}.  The criterion for estimating the linkage between the clusters is based on the average distance between pairs of  signals among the clusters, weighted by the numbers of elements in each cluster. Cluster linkage at each level of dendrogram is calculated recursively whose value for a given pair of clusters is given by 

\begin{equation}
d(r,s) = \sqrt{\frac{2n_{r}n_{s}}{(n_{r}+n_{s})}}  \| \tilde{x}_{r} - \tilde{x}_{s} \|_{2}
\end{equation}

The  optimal distance measure used for linkage  and the  original mother wavelet used for decomposition are both selected based on the  value of cophenetic correlation coefficient,c  \cite{sneath1973numerical} with value close to unity being ideal. 

\begin{equation}
c = \frac{\sum\limits_{i<j}(Y_{ij} - y)(Z_{ij} - z)}{\sqrt{\sum\limits_{i<j}(Y_{ij}-y)^{2}\sum\limits_{i<j}(Z_{ij}-z)^{2}}}
\end{equation}

$Y_{ij} $ is the pairwise distance between parametrized waveforms while $Z_{ij}$ is their linkage distance. $y, z$ respectively represent the average value of the corresponding distance measures. 

Optimal leaf ordering of the resulting dendrogram is achieved by maximising the sum of similarities between adjacent leaves \cite{Bar-Joseph01062001}.  
This step is carried out to identify the relationship between the various clusters and to locate possible subgroups. For example in Figure \ref{fig:Dendrogram}, transients at both  ends are least related to each other. The schematic of the hybrid classifier useful for real time transient classification is shown in Figure \ref{fig:Schematic}.

\begin{figure*}[t]
\includegraphics[width=0.475\textwidth]{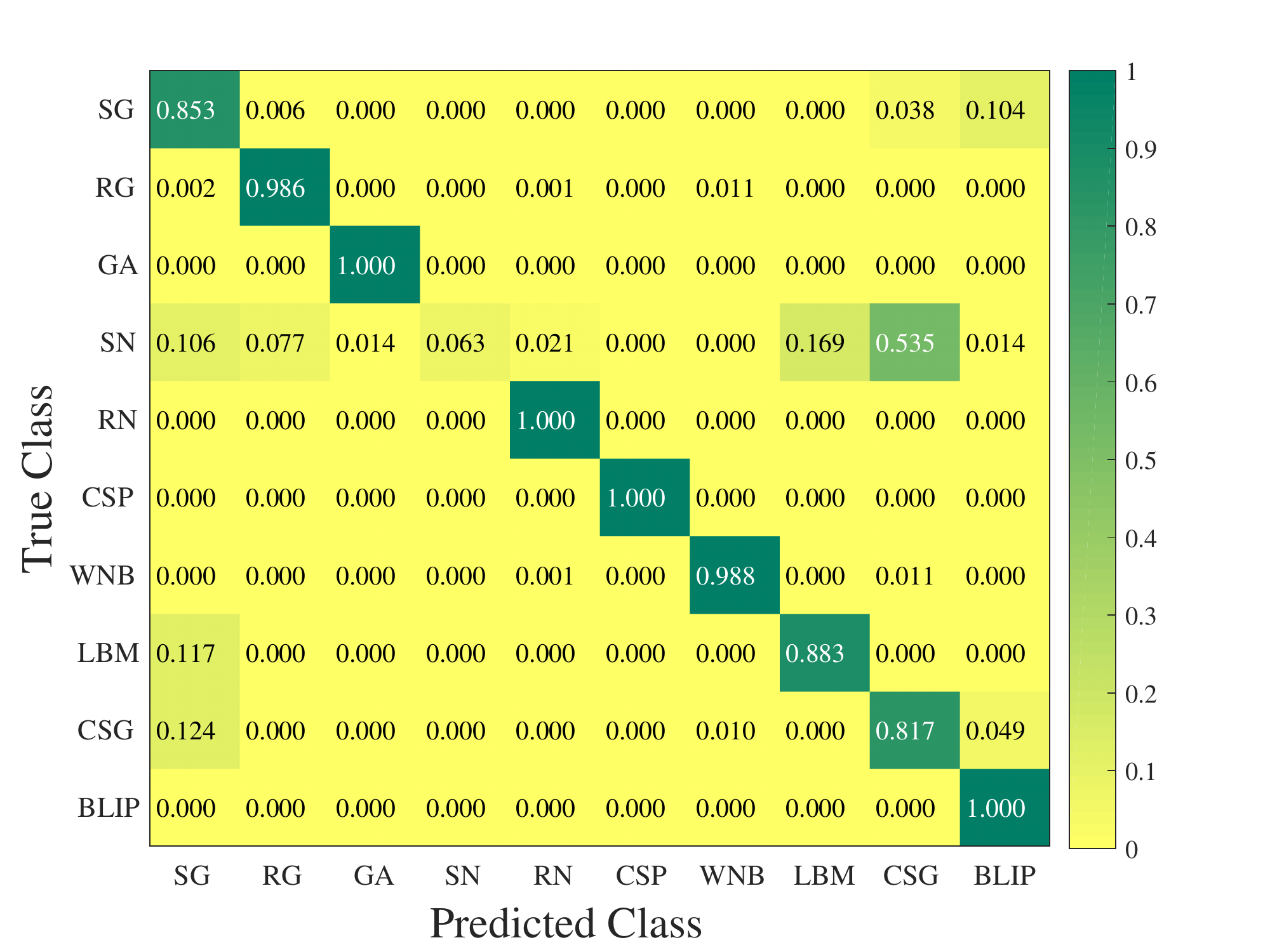}
\includegraphics[width=0.475\textwidth]{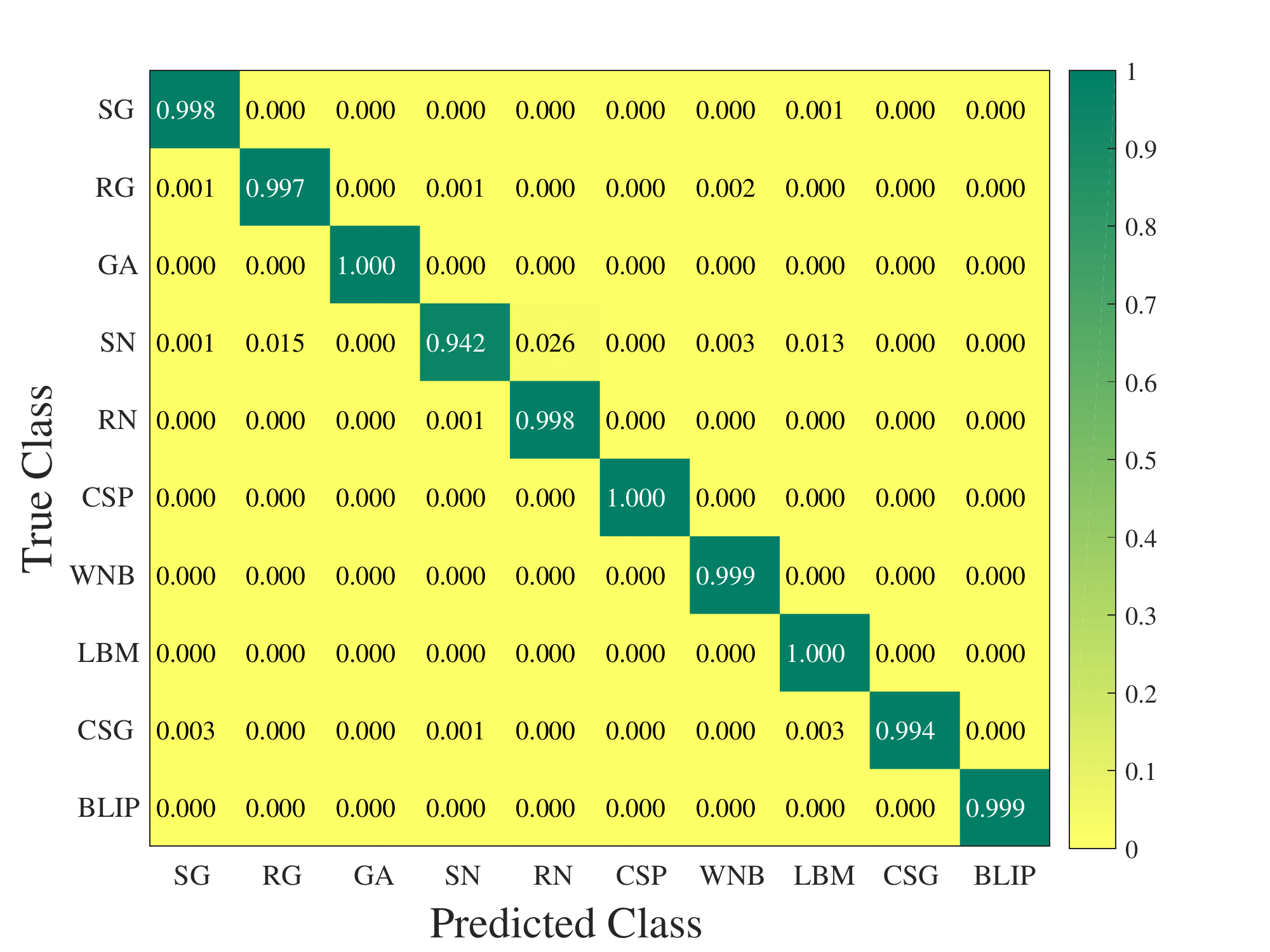}
\caption{\label{fig:simDBNN_test} Confusion Matrix for Simulated Data:  Results from traditional SVM (left) and DBNN (right) classifiers. Closer a diagonal element to unity, better is the classification for the corresponding type. Accuracy of our method is thus evident.}
\end{figure*}

\begin{table}[!htb]
\caption{\label{tab:glitches}Details of transients used. A=Astrophysical, NA=Non-Astrophysical, O1: Advanced LIGO 1st Science Run, S6=LIGO 6th Science Run, Sim=Simulated }
\begin{tabular}{lcccc}
\hline\hline
Transient  & Symbol & Type & Search Type  \\
\hline 
Sine Gaussian \cite{abbott2004first, 0264-9381-31-1-015016} & SG & NA & S6, Sim    \\
Ring Down \cite{vishveshwara1970scattering}  & RD  & A & S6, Sim  \\ 
Gaussian \cite{abbott2004first, PhysRevD.85.122007}  & GA & NA & S6, Sim    \\
Supernova  \cite{Zwerger:1997sq,Burrows:2005dv} &  SN & A & S6, Sim   \\
Cusp \cite{PhysRevLett.85.3761} & CSP &  A & S6, Sim   \\
White Burst Noise \cite{PhysRevD.85.122007} &  WNB & NA & S6, Sim   \\
Black Hole Merger \cite{PhysRevD.65.044001,PhysRevD.65.124012} & LBM & A &S6, Sim   \\
Chirping Sine Gaussian \cite{2016arXiv160606096B}  & CSG  & NA & Sim  \\ 
Blip \cite{abbott2016characterization}  & Blip & NA & Sim    \\
Scattering \cite{abbott2016characterization,accadia2010noise, ottaway2012impact}  &  SCT & NA & O1  \\
Type A (Low Frequency) &  A & NA& O1  \\
Type B &  B & NA& O1  \\
Type C (Blip, Fig~\ref{fig:OmegaScan}) \cite{CQG_Article}  &  C & NA& O1  \\
Type D  &  D & NA& O1  \\
Type E (Koi Fish) \cite{CQG_Article} &  E & NA& O1  \\
Type F (Needle) &  F & NA& O1  \\
Type G &  G & NA& O1  \\
Lightning \cite{abbott2016characterization}&  LGN & NA & Targeted \\
\hline\hline
\end{tabular}
\end{table}

\section{\label{sec:Wavelet} Feature Extraction}
Raw time series is preprocessed by applying a whitening transformation which enhances the short duration features seen in data. Transient signals occurring in power systems and neuro magnetic brain responses have structural and temporal similarities with the glitch signals found in LIGO data streams. Wavelet based feature extraction for classifying these transients are detailed in \citep{bhagat2009investigating, jayasree2009classification}.  Wavelet is a function having a smooth oscillatory pattern which vanishes near the ends \citep{bhagat2009investigating}. With its desirable qualities like good localization in time and frequency domain, it seems to be a natural choice for extracting information from transient signals.  Discrete wavelet transform results in sparser signal representation consisting of reduced feature set, but still preserves information necessary to differentiate among the classes. 
\\

The mother wavelet is defined as, 
\begin{equation}
\psi_{a,b}(t) = |a|^{-1/2}\psi(\frac{t-b}{a})   
\end{equation}
where a,b $\in$ R and a $\neq$ 0  
is scaled and time-shifted to form the wavelet family. 

Orthonormal basis of Hilbert space $L^{2}(R)$ consisting of finite-energy signals is obtained by discretising  scale and translation parameters $a_{j} = 2^{-j}$ and $b_{j,k} = 2^{-j} k$ giving the family wavelet as 

\begin{equation}
\psi_{j,k}(t) = 2^{j/2}\psi (2^{j}t - k)  \;with\; j,k \in Z
\end{equation}

The all resolution level wavelet decomposition of the signal has form:

\begin{equation}
S(t) = \sum_{j = -N}^{-1} \sum_{k}C_{j}(k)\psi{j,k}(t) = \sum_{j = -N}^{-1} r_{j}(t)
\end{equation}

where N = log2 (signal length). 

The energy $E_{j}$ at each resolution level is computed as:
  
\begin{equation}
E_{j} = \|r_{j}\|^{2} = \sum_{k} |C_{j}(k)|^{2}
\end{equation}

The relative wavelet energy at each resolution level 

\begin{equation}
p_{j} = \frac{E_{j}}{E_{tot}}, \quad \textrm{where} \quad E_{tot} = \sum_{j<0}E_{j}
\end{equation}

Wavelet entropy $S_{WE}$ which encodes the degree of disorder in a signal can be written as,

\begin{equation}
S_{WE} = -\sum_{j<0}p_{j} \;ln[p_{j}]
\end{equation}


Here we carry out similar N-level one-dimensional wavelet decomposition \cite{MATLAB_2013} using an appropriately chosen mother wavelet. Feature extraction for simulated and LIGO O1 data is done using Daubechies 2 ({\tt db2} ) wavelet  while for other search cases discrete Meyer ({\tt dmey}) wavelet is used. We use $N=12$ and $14$  respectively for data sampled at 4 and 16 KHz.  Energy in the detail levels and wavelet entropy are then computed and are normalized to unity. In addition, kurtosis of the whitened signal is also used as a distinguishing feature. These features along with the class labels form the input for our Bayesian classifier. Figure~\ref{fig:ip} shows typical transients and their detail coefficient wavelet energy.

\section{Simulated Data}
Simulated data set consists of 49845 transients from 10 classes (refer Table ~\ref{tab:glitches}) whose SNR is varied uniformly  between 8 and 100 by means of Gaussian white noise addition. Signals in each class are generated for different values of  parameters sampled from a wide range. Details of bursts used in simulation are given below ($t _{o}$ is set to 0.5 Sec ).

\begin{itemize}

\item{Gaussian (GN)}

These broadband non-astrophysical signals are modelled as simple Gaussians with duration parameter $\tau$ taking values 0.0005, 0.001, 0.0025, 0.005, 0.0075, 0.01, 0.02 and 0.05 \cite{abbott2004first}.
\begin{equation*}
s(t) =   exp(- \frac{(t-t_{o})^2}{\tau^2})
\end{equation*}

\item{Sine-Gaussian (SG)}

SG models a non-astrophysical glitch which produces significant triggers in matched  filtering analysis for coalescing compact binaries \cite{0264-9381-31-1-015016}. $\tau$ is set to $2/f_{o}$ with central frequency ($f_{o}$)  logarithmically spanning from 100 Hz to 2000 Hz.
\begin{equation*}
s(t) =   exp(- \frac{(t-t_{o})^2}{\tau^2}) \; \sin (2 \pi f_{o} (t-t_{o}) )
\end{equation*}

\item{Ringdown (RG)}

RG signals have longer duration but shorter bandwidth and are modelled as damped sinusoids. They are produced from quasi-normal modes of a final black hole formed from coalescing compact binaries \cite{vishveshwara1970scattering}. Here we set $\tau = 4/f_{o}$ with $f_{o}$ similar to that of Sine-Gaussian data set.
\begin{equation}
 s(t) =
  \begin{cases}
    s(t) = exp(- \frac{(t-t_{o})}{\tau}) \; \cos (2 \pi f_{o} (t-t_{o}))     & \quad \text{if } t \geq t_{o} \\
    0  & \quad \text{if } t \leq t_{o} \\
  \end{cases}
  \end{equation}
  
\item{Chirping Sine Gaussian (CSG)}

CSG is similar to SGs but with an additional chirping parameter \cite{2016arXiv160606096B}. This signal closely models the whistle glitches frequently seen in LIGO detector data. Equation below gives the waveform model where each of the parameter is varied as follows:  $f_{o}$:\{5,100\}, $\alpha$:\{10,100\} and $\tau$:\{0.001,0.025\}
\begin{equation*}
s(t) = \frac{\exp{ \bigg{(}\frac{-(1-1i\alpha)\;(t-t_{o})^{2}}{(4\tau^{2})} + 2\pi i(t-t_{o})f_{o} }\bigg{)}}{(2\pi\tau^{2})^{\frac{1}{4}}}\end{equation*}

\item{Supernova (SN)}

Zwerger-Mueller  waveforms \cite{Zwerger:1997sq} , one of the Supernova waveforms, are produced by axi-symmetric core collapse of supernovae. These are obtained by hydrodynamical simulations of stellar core collapse by varying the  initial conditions like adiabatic index, spin, and differential rotation profile. We incorporate 78 models (with varying SNR) consisting of a simple analytic equation of state. We also make use of Ott-Burrows supernova waveforms \cite{Burrows:2005dv} in our analysis.

\item{Cusp (CSP)}

Symmetry breaking phase transitions in early universe could generate cosmic strings \cite{PhysRevLett.85.3761}  with a cusp like signal, $h(f) = A(f) f^{-4/3}$. Such waveforms are simulated with exponential roll off after a cut-off frequency $f_{o}$ which is varied from 50 Hz to 2000 Hz. 

\item{White Noise Bursts (WBN)}

WBN have in general very complex time-frequency morphology.  Their spectra is white in the specified band and zero outside\cite{PhysRevD.85.122007}. Here we construct a set of  burst signals  which have central frequency  spanning from 50 - 300 Hz, bandwidth 50 to 150 Hz and duration $0.1$ to $0.4$ seconds.

\item{Black Hole Merger (LBM)}

These waveforms capture the coalescence radiation emitted from a merger of binary black-hole systems using Lazarus approach \cite{PhysRevD.65.044001}. Analytic approximation  \cite{PhysRevD.65.124012} is used to construct time domain templates to replicate the merger scenarios. We considered black hole binaries with a chirp mass in range $\{20,50\}$ and cosine of inclination angle varied between zero and one.

\item{Blip (Blip)}

Blips are observed frequently in both LIGO detectors but their origin is not well understood \cite{CQG_Article}. Hardware injections carried out at the observatories sometimes hit the saturation limit of the actuator resulting in signals which look similar to blips. Hence we simulate them by clipping Sine-Gaussians at few percent level around the mean amplitude.

\end{itemize}

While sampling the parameters, care was taken to ensure that the signals within a class are significantly different. Table~\ref{tab:sim_test} shows the performance of Wavelet-DBNN Classifier. Total number of samples, size of training set, true positives (TP), false positives (FP), precision, sensitivity and specificity are reported (see \cite{powers2007evaluation} for terms definition). The resulting confusion matrix is shown on the right panel of Figure~\ref{fig:simDBNN_test}.

\begin{table}[h]
\caption{\label{tab:sim_test} Simulated transient signals.}
\begin{tabular}{lccccccc}
\hline\hline
Name &Total & Train. & TP & FP & Preci. & Sensi. & Speci. \\
\hline
SG  & 5000 & 552 &4991  &   22  &  0.99  & 0.99  & 1.00 \\
RG  &  5000 & 311 &4984  &  16 &  0.99  & 0.99  & 1.00 \\
GA  & 5000 & 155 &5000 &   0 &  1.00 &  1.00 & 1.00 \\
SN  & 745 & 313 &  702   &  14  & 0.98  & 0.94  & 1.00 \\
RN  & 5000 & 114 &4992  & 19  & 0.99 & 0.99  & 1.00 \\
CSP  &5000 & 14 & 5000  & 0  &  1.00  & 1.00  & 1.00 \\ 
WNB &  10000 & 421 & 9994  & 15  & 0.99  & 0.99  & 1.00\\
LBM  & 5000& 586 &4999 &  31 & 0.99 &  1.00  & 0.99\\
CSG  & 5000& 378 &4969 &  0 & 1.00  & 0.99  & 1.00 \\
BLIP  & 4100& 153 & 4097  & 0 & 1.00  & 0.99 &  1.00 \\
\hline\hline
\end{tabular}
\end{table}

For comparison with a standard classifier, we use publicly available support vector machine (SVM) implementation {\tt LIBSVM} \cite{ cc01a, Octave} on the same wavelet decomposed parameter sets. Figure~\ref{fig:simDBNN_test} clearly shows how our Wavelet-DBNN classifier outperforms the traditional classifier. Stark difference is observed for Supernova signals where the SVM  shows very high misclassification, most likely due to the limited number of data samples and the inherent diversity in their morphology. 

\section{S6 hardware Injections}
To check the performance of our classifier on the real data, we use the classifier on the LIGO strain data obtained from the sixth science run \cite{0264-9381-32-11-115012}.  We apply our classifier to six different classes of hardware injected short duration transient signals as given in Table~\ref{tab:real_test}. The strain data is whitened to better identify the transients and then down sampled to 4096 Hz. 1634 transients with SNR greater than 10 are used in the analysis.  Table~\ref{tab:real_test} gives the results after classification.  

\begin{table}[h]
\caption{\label{tab:real_test} S6 Hardware Injections.}
\begin{tabular}{lccccccc}
\hline\hline
Name & Total & Train. & TP & FP & Preci. & Sensi. & Speci \\
\hline
SG  & 1476 & 69 &   1476 & 10 & 0.99 & 1.00 & 0.94 \\
RG  & 36  & 25 &  33 & 0 & 1.00 & 0.92 & 1.00 \\
GA  & 46 & 38 &  44 & 8 & 0.85 & 0.96 & 1.00  \\
SN  &  41 & 34 &  33 & 4 &  1.00 & 0.86 & 1.00  \\
CSP  & 28 &  27 &  24 & 0 & 1.00 & 0.86 & 1.00 \\
WNB &  29 & 27 &  24 & 0 & 1.00 & 0.83 & 1.00\\
\hline\hline
\end{tabular}
\end{table}

\section{ Targeted Search: LIGO Strain Channel}

\begin{figure}[!htb]
\includegraphics[width=0.475\textwidth]{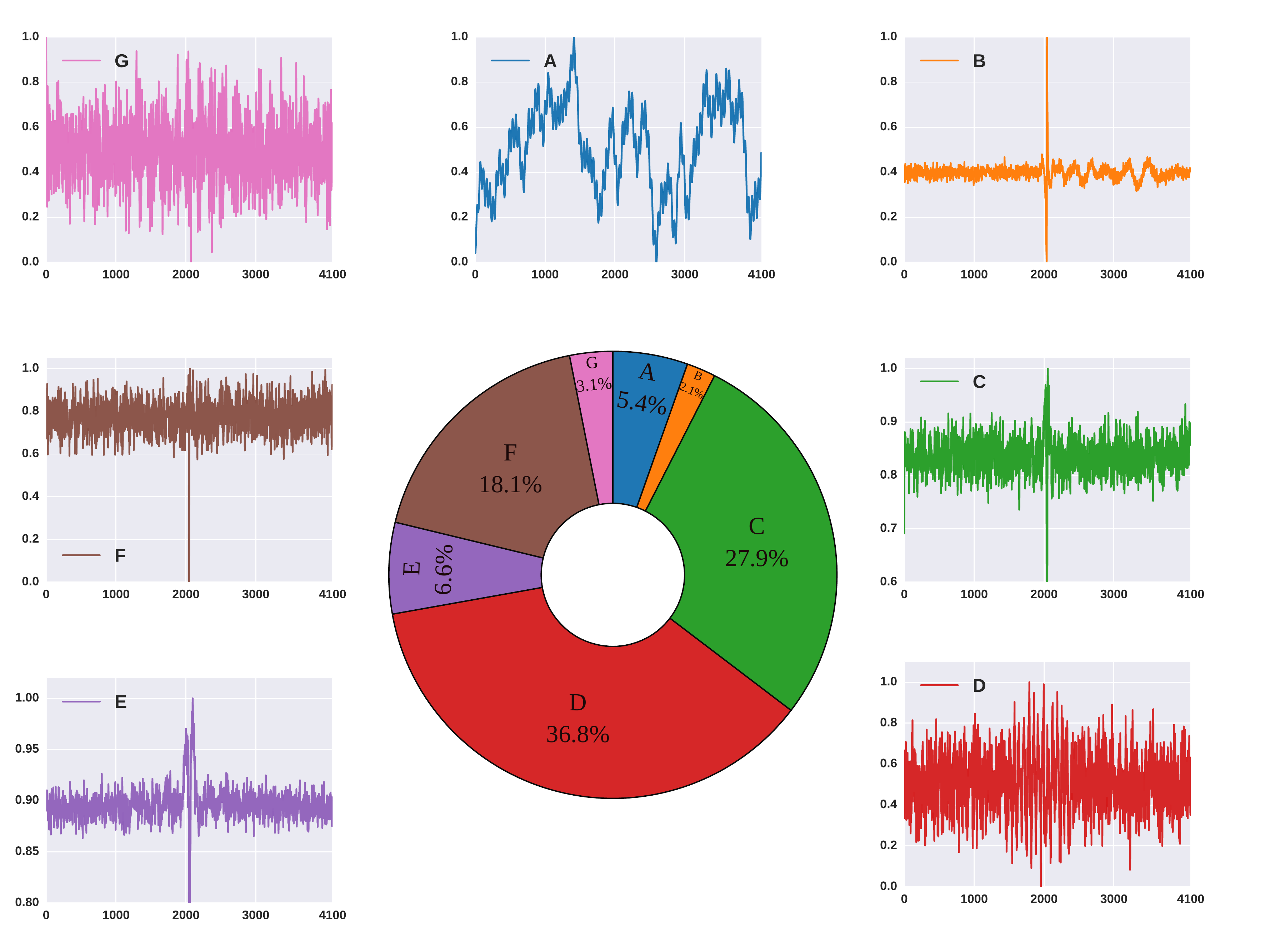}
\caption{Donut chart shows the distribution of O1 transients classified by DBNN-Wavelet classifier into the major classes identified by Hierarchical Clustering. Representative transients from each class  (sampled at 4096 Hz)  are also shown. }
\label{fig:O1_Pie}
\end{figure}

Detector Characterization studies revealed several kinds of non-astrophysical transients in the Advanced LIGO's strain data during its first observation run ~\cite{abbott2016characterization}. Identification of these transients and establishment of their non-astrophysical origin were crucial for the detection of GW signal \cite{GW150914,GW151226,abbott2016characterization}. For those known classes, if we could automate their detection using machine learning methods, it will reduce the noise background in the astrophysical GW searches. We used LIGO Hanford Observatory (LHO) strain data (September 18 to January 12) consisting of 28354  transient triggers~\cite{OMICRON}  with SNR ranging from 8 to 100 and having a maximum frequency of 2096 Hz. 
 Our classifier is then used to search for events which look similar to the major transient classes (see Table~\ref{tab:glitches})  evident from the initial hierarchical clustering. Several of these could potentially limit generic burst searches in particular Cosmic Cusps and Supernova events. One second whitened data around the trigger is used for feature extraction. Training set consisted of minimal samples ranging from five to ten per class. Figure~\ref{fig:O1_Pie} shows the distribution of classified  transients with similar morphology as the training set.
 
We also did the analysis with LIGO Livingston data for triggers in SNR range $10$ - $100$ and obtained comparable results. Classifier found a strong presence of glitches caused due to previously proposed scattering mechanism ~\cite{accadia2010noise, ottaway2012impact}.
We observe coincidence (within  one second) between some of the classified scattering glitches and the triggers seen in LIGO's auxiliary angular length sensing channels. These auxiliary channels carry information about the motion of signal recycling cavity optics. The coupling was seen to occur predominantly from the pitch and yaw degrees of freedom with a respective contribution of 17.3\% and 12.5\% with 50\% of the glitches coincidently seen in both the channels. Scattering happens when off-axis beam gets reflected back from the beam tube and recombines with the main beam.  These morphology  based identification coupled with coincident analysis would help one to narrow down to the region mostly likely to cause the transients and also help in appyling appropriate data quality vetoes. 

\section{ Targeted Search: LIGO Auxiliary Channels}

\begin{figure}[!htb]
\includegraphics[width=0.475\textwidth]{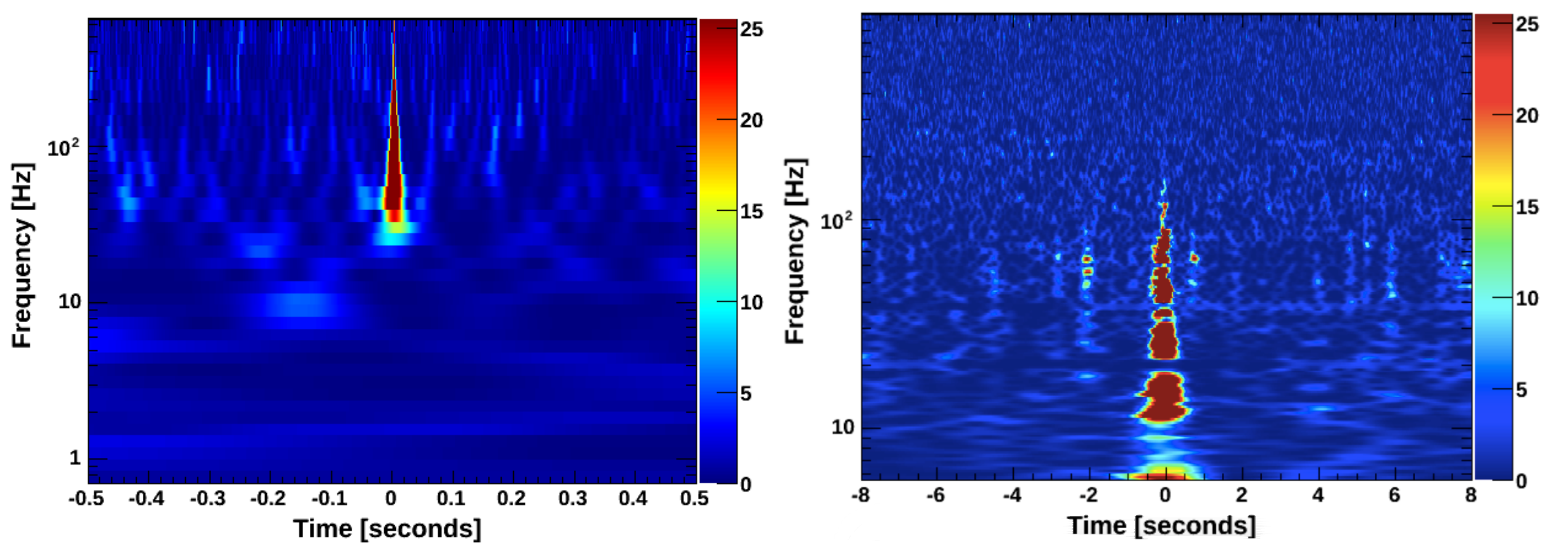}
\caption{OmegaScan: Type C glitch in LHO strain channel (left) and lightning glitch in LLO magnetometer (right). Plots generated through LIGOdv-web \cite{LigoDV}.}
\label{fig:OmegaScan}
\end{figure}

Severe weather conditions can affect both the detectors and, if not properly vetoed, can be misinterpreted as a true signal. Variation in the ambient magnetic field during lightning and thunderstorm around LIGO can affect sensors and actuators present in multi-stage suspension systems that isolate and control the LIGO test mass. They are seen in magnetometers with a very distinct time-frequency morphology (Figure~\ref{fig:OmegaScan}, right panel). These also induce currents in the beam tube  and are detected by on site clamp meters.  Here we apply our classifier to separate out lightning events from the other transients seen in the magnetometer data. We use  LIGO Livingston Observatory (LLO) Y-arm magnetometer Omicron triggers  generated from 16:00:00 to 23:00:00 UTC of December 16, 2015. Hierarchical clustering on the first 30 minutes of data generates the training set which is fed to the supervised classifier that performs the final targeted search.  
Triggers with SNR 15 to 1000 and frequency 1 to 1024 Hz are used for the analysis. 42 out of 689 such triggers are identified to be caused by lightning. Similar search carried out in LLO X-arm magnetometer data for the same period identifies 45 lightning triggers. Our results are consistent with the local weather data which reported lightning activity during the same period.  Number of misclassifications in these cases turned out to be only 1 and 6 respectively.

\section{Conclusions}

We have convincingly demonstrated the resourcefulness of machine learning in detector characterization and burst signal analysis in LIGO like complex instruments. We showed that an effective feature extraction technique, in conjunction with an efficient classifier, can be used to classify a variety of transients in practical situation involving real data. We used relative wavelet energy, wavelet entropy and kurtosis as a possible parameter set for classifier input. This, coupled with a difference boosting neural network, was very accurate in discerning between classes with slightly different morphology and possibly different physical origin. The usefulness of the method was shown in our analysis where we could do an accurate targeted search for a specific glitch using minimal training sets. The parameter set used here can be expanded to include other features which can aid the classification even when the corresponding values are unavailable for other classes. The special construction of the classifier makes sure that it does not suffer from the curse of dimensionality unlike most neural network classifiers. Hence the feature set can be expanded in future without causing much computational overhead.  Combining class information along with multi-channel coincidence analysis will help to narrow down to the cause for a particular kind of transient present in the data. If there is good enough reason to believe that the trigger is non-astrophysical then glitch based vetoes can be applied to those times.  This would lower background triggers in search pipelines thus enhancing confidence in the true detections.  We plan to develop such a data quality vector which can be used to directly veto low latency triggers produced by search pipelines looking for astrophysical signals.

\section{Acknowledgements}

We would like to thank the Detector characterization working group of the LIGO Scientific Collaboration for useful comments and suggestions. NM acknowledges Council for Scientific and Industrial Research (CSIR), India for providing financial support as Senior Research Fellow. SM acknowledges the support of the Science and Engineering Research Board (SERB), India through the fast track grant SR/FTP/PS-030/2012. Authors express thanks to  Siddhartha Chatterjee, Arun Aniyan, Shantanu Desai and Bhooshan Gadre for their valuable comments and suggestions. LIGO was constructed by the California Institute of Technology and Massachusetts Institute of Technology with funding from the National Science Foundation and operates under cooperative agreement PHY-0757058. This paper has been assigned LIGO document number LIGO-P1600094.

\label{Bibliography}
\bibliography{GWB_Classification}

\end{document}